\documentclass[aps,prl,twocolumn,noshowpacs,superscriptaddress,floatfix]{revtex4}

\usepackage{amsmath}
\usepackage{amssymb}
\usepackage{graphicx}
\usepackage{dcolumn}
\usepackage{bm}
\usepackage{color}
\usepackage{ulem}
\usepackage{booktabs}

\begin{document}

\title{Network structure of thermonuclear reactions in nuclear landscape}

\author{H. L. Liu}
\affiliation{Key Laboratory of Nuclear Physics and Ion-beam Application (MOE), Institute of Modern Physics, Fudan University, Shanghai 200433, China}
\affiliation{Shanghai Institute of Applied Physics, Chinese Academy of Sciences, Shanghai 201800, China}
\affiliation{University of the Chinese Academy of Sciences, Beijing 100080, China}

\author{D. D. Han}\thanks{ddhan@fudan.edu.cn}
\affiliation{School of Information Science and Technology, Fudan University, Shanghai 200433, China}

\author{Y. G. Ma}\thanks{mayugang@fudan.edu.cn}
\affiliation{Key Laboratory of Nuclear Physics and Ion-beam Application (MOE), Institute of Modern Physics, Fudan University, Shanghai 200433, China}
\affiliation{Shanghai Institute of Applied Physics, Chinese Academy of Sciences, Shanghai 201800, China}

\author{L. Zhu}
\affiliation{Shanghai Institute of Applied Physics, Chinese Academy of Sciences, Shanghai 201800, China}
\affiliation{University of the Chinese Academy of Sciences, Beijing 100080, China}

\date{\today}

\begin{abstract}
Nucleosynthesis is a complex process in astro-nuclear evolution.  In this work, we construct  a directed multi-layer nuclear reaction network using  the substrate-product method from a thermonuclear reaction database, JINA REACLIB. The network  contains four layers, namely $n$-, $p$-, $h$- and $r$, corresponding to the reaction types involved in neutrons, protons, $^4$He and the remainder, respectively. The degree values (i.e. numbers of reactions)  for three layers of n-, p- and h- have a significant correlation with one another, and their topological structures exhibit a similar regularity. However, the $r$-layer has a more complex topological structure than others and has less correlation with the other three layers. A software package named `mfinder' is employed to analyze the motif structure of the nuclear reaction network. We thus identify the most frequent reaction patterns of interconnections occurring among different nuclides. This work provides a novel approach to study the nuclear reaction network prevailing in the astrophysical context.

\end{abstract}

      

\maketitle


\section{1. Introduction} 

 Nucleosynthesis from light particles into heavier ones in the early universe and the corresponding nuclear processes are of significant recent interest in the nuclear astrophysics \cite{burbidge1957,schatz1998,arnould2003,arnould2007,kappeler2011,Chen,ji2016,Fynbo,An2,An1,NST_Pais,NST_Tang,NST_Ma,NST_Li,LiZ2}. 
Much attention has been paid earlier on topics like the exploration up to limits 
{of the nuclear landscape} and searching for the island of stability \cite{erler2012,WangR,SHE,SHE2,NST_Yu,NST_Daderi,NST_Boilley,Qian}.  {The terrestrial nuclear reaction experiments provide a large volume of data for thermonuclear reaction rates (mainly on stable targets), based on which many theoretical models have been developed to calculate the binding energy of different nuclides and the cross-sections and reaction rates for different nuclear processes} \cite{Herndl,Ota,CPC1,CPC2,WangPLB,WangEPJA,WangK,XuXX,MaPLB,LiZH,Jiang,Yun,Wei,Wu,Ben,NST_Guo,NST_Liu,NSTT,Yancin}.  The JINA REACLIB database is a well-known source of thermonuclear reaction rates ~\cite{CH2010-22}, {updated continuously and snapshotted regularly by the JINA Collaboration which aims to compile a complete set of nuclear reaction rates.} The dataset contains various kinds of nuclear reactions, parameters for calculating reaction rates, $Q$ values, \textit{etc.}, providing rich information for nuclear-related research directions. 

Traditionally, calculations of nuclide abundance {can be} realized by solving sets of time-dependent differential equations in the database, however, it is rather difficult and  {challenging.
An alternative method has been proposed by a  complex network construction to explore its {topological} characteristics of thermonuclear reactions  in our previous work \cite{ZL2016-23}.
{The scope of the complex network theory is also extended to the social system and chemical reaction processes}~\cite{WS1998-16,JCC2010-17,JCC2012-18}}. {The main idea is to identify the real interacting units as nodes and relationships between two units as edges, which form the basis of the graphical structure of a network.
By doing so, translation to networks can thus enable identification of topological characteristics and other hidden features of the real systems. For example, there exists a self-similar `scale-free' structure with  power-law degree distribution  emerging in many systems, eg. the Internet network, naturally occurring objects like biological cells and tissues~\cite{BA1999-19,BS2006-20,CL2011-21}. Other real-world systems having Poisson degree distribution, such as road networks, also exist  \cite{road}.}

\begin{figure}
\includegraphics[width=8.5cm]{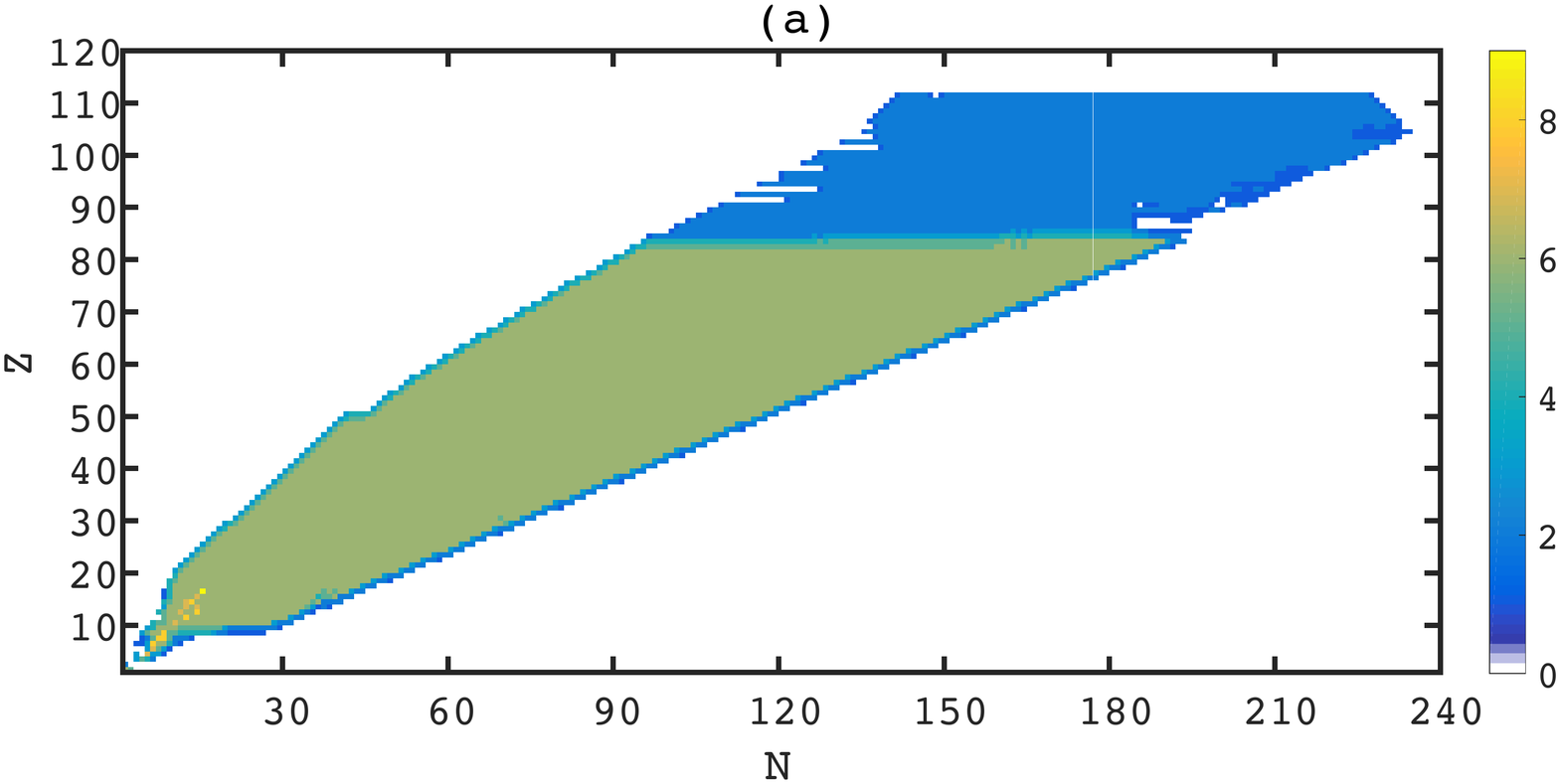}
\includegraphics[width=8.5cm]{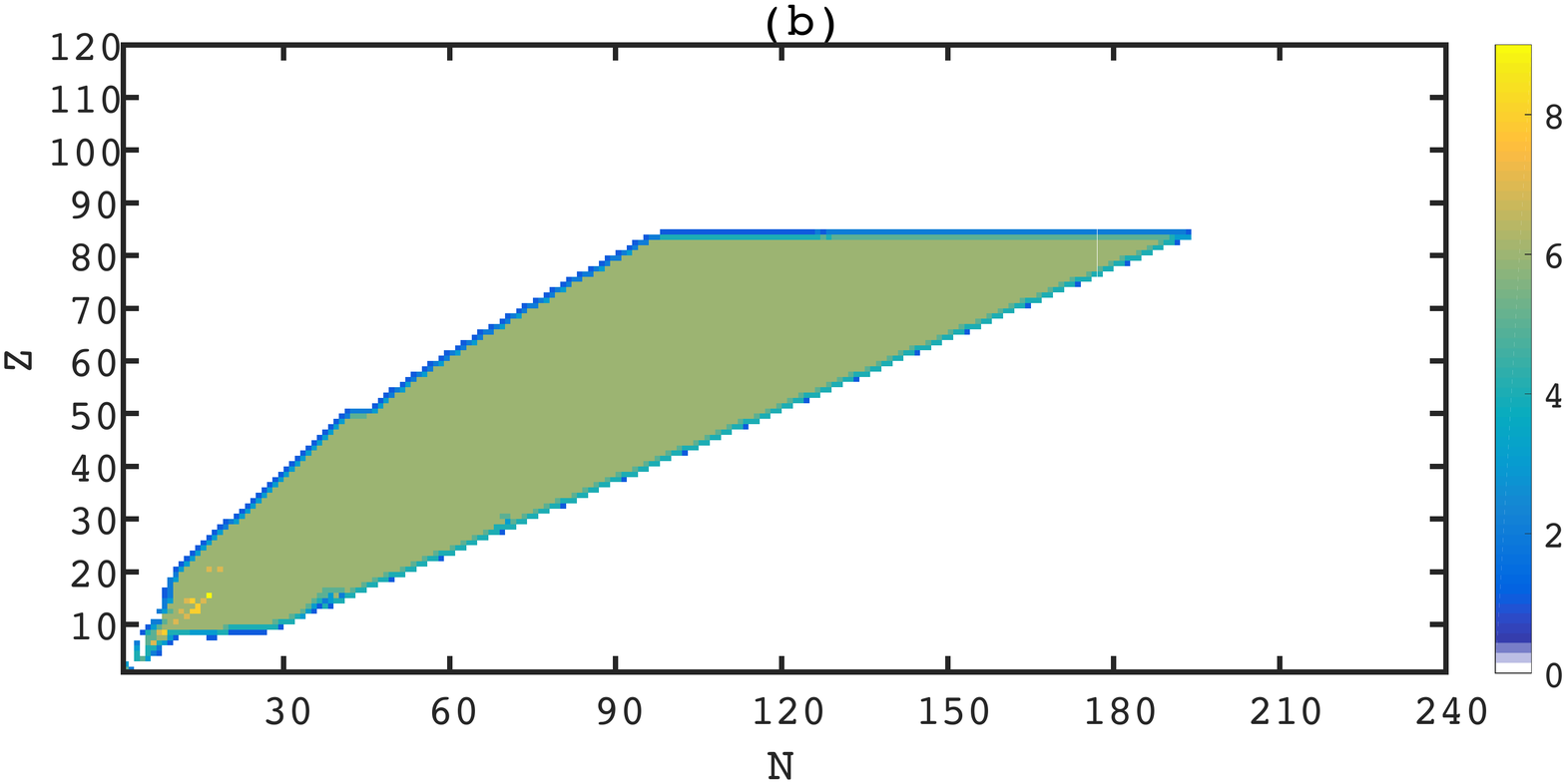}
\includegraphics[width=8.5cm]{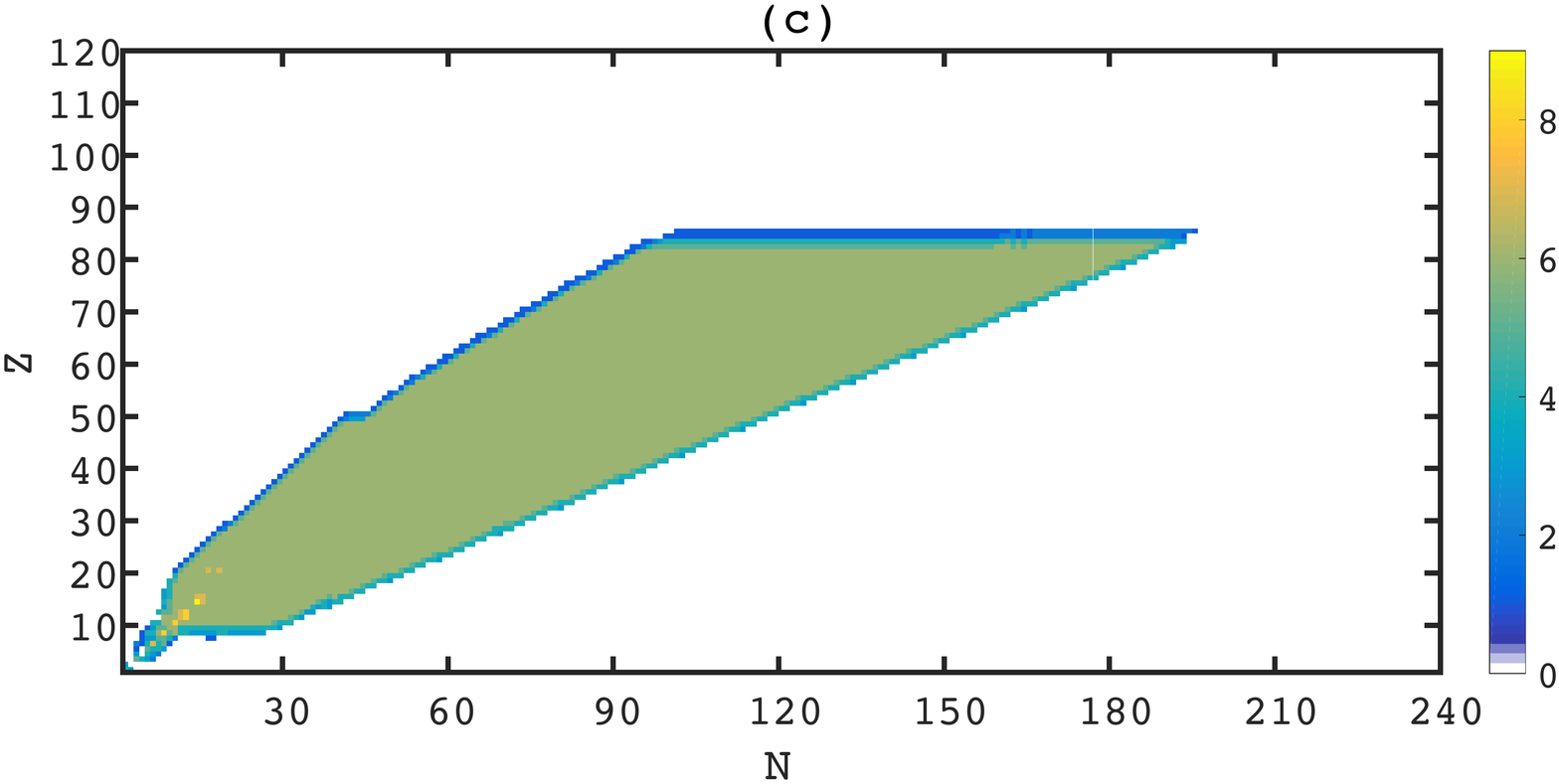}
\includegraphics[width=8.5cm]{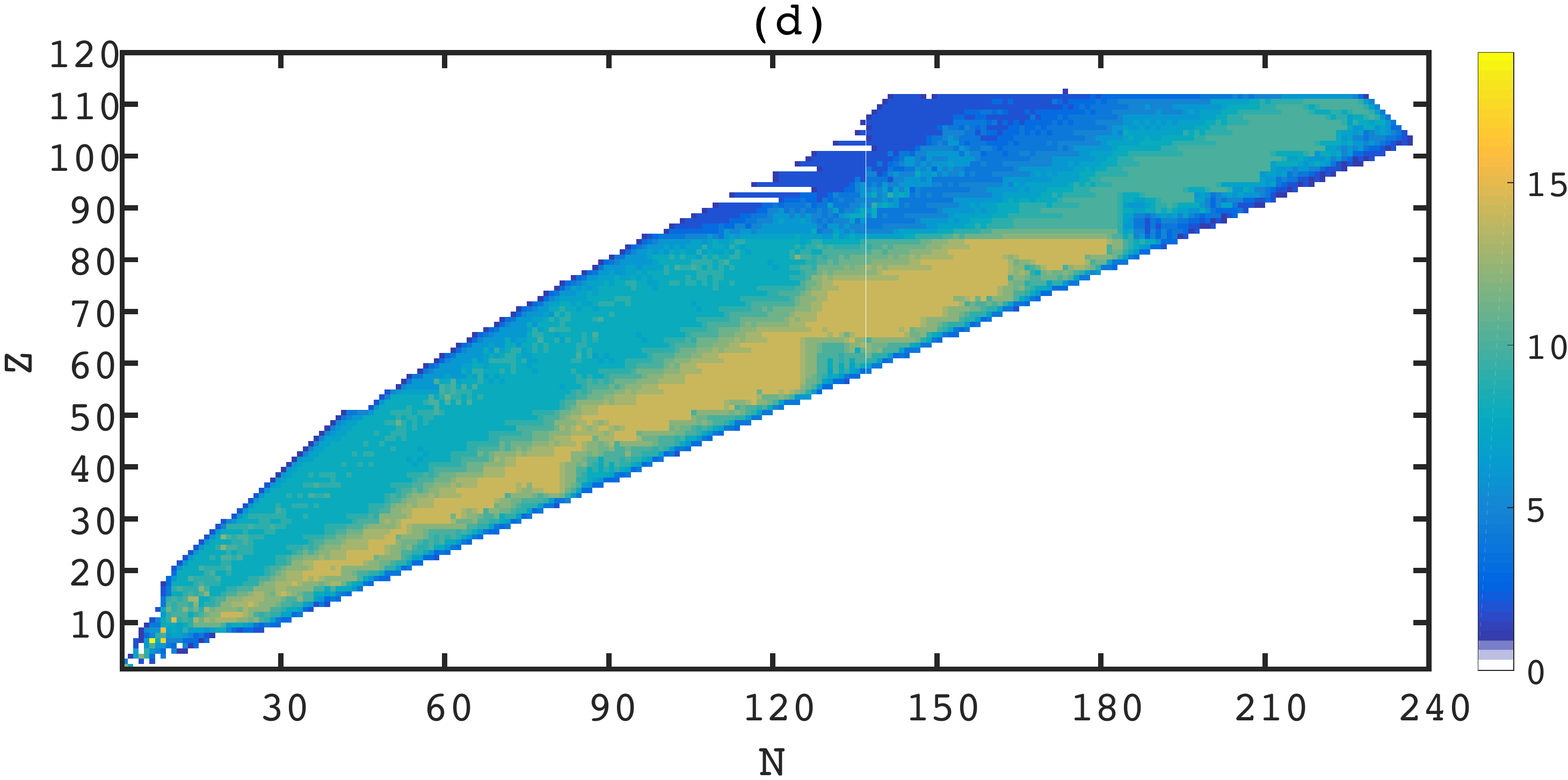}
\caption{The degree value of each nucleus in $n$-layer (a), $p$-layer (b), $h$-layer (c) and $r$-layer (d) plotted in the $N-Z$ plane. The different colors represent different degree values.}
\label{degree}
\end{figure}

In our previous work ~\cite{ZL2016-23}, {for the first time} a complex network construction for { the }nuclear reaction had been framed {exploring different} nuclei and reactions based on  JINA REACLIB database~\cite{CH2010-22}.
This complex network is a {class} of multi-layer  nuclear reaction networks where { the constituent four layers namely, $n$, $p$, $h$, and $r$ which are labeled according to each linked edge with four types of particles, i.e. neutron, proton, $^4$He and the remainder.} It is found that { the} nuclear reaction network exhibits bimodal degree distributions, which is very different from  either power-law or Poisson distributions. Combining topological degree value{s} of the nuclei along the stability line, it was found that over 80$\%$ of the stable nuclides follow the condition $(K_{i,\mathrm{in}}^{[r]}-K_{i,\mathrm{out}}^{[r]}=2)\bigcap(K_{i,\mathrm{in}}-K_{i,\mathrm{out}}=0)$, where $K_{i,in}^{[r]}$, $K_{i,out}^{[r]}$, $K_{i,in}$ and $K_{i,out}$ {present the in-coming and out-going degree value of the nuclei} in the $r$-layer and aggregated network, respectively~\cite{ZL2016-23}. {The novelty in the suggested method lies in that it can address a stable  nucleus with a half-life lower than $10^9$ years (in contrast with the traditional method)}~\cite{MT2004-24}.

Based on {our} success to describe network properties in the nuclear landscape {previously}, here we use {a similar} multi-layer nuclear reaction {network} to explore {their topological structure and find  the properties of motifs featuring nuclear reaction patterns for nuclides in different layers}. The degree correlation{s} between different layers are also investigated.{ The present work especially elucidates some novel features of nuclear reactions prevailing in nuclear astrophysics.}

\begin{figure}
\vspace{-0.5cm}
\includegraphics[width=9cm]{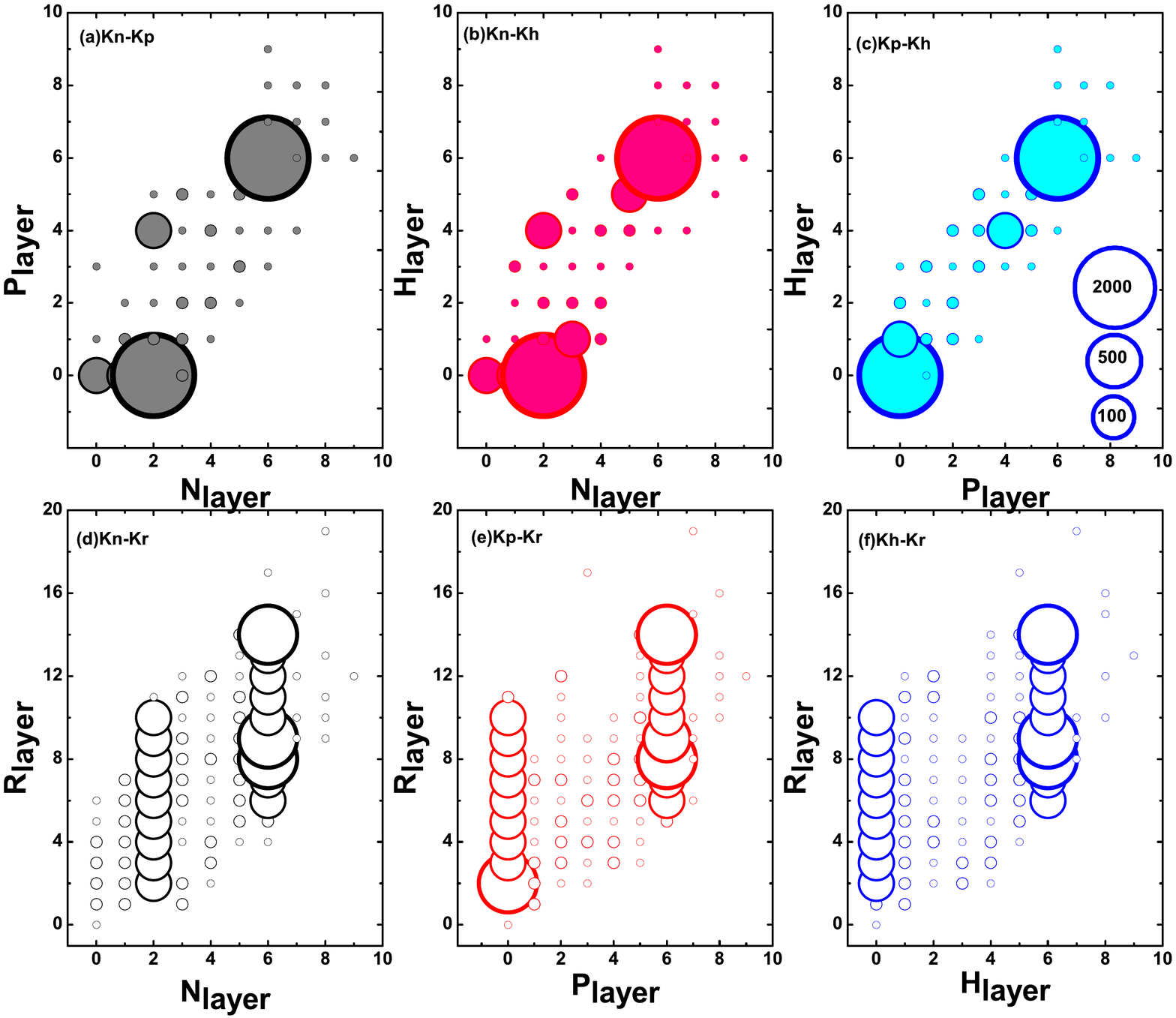}
\vspace{-0.6cm}
\caption{Degree-degree correlation for each two layers  among $n$-, $p$-, $h$- and $r$-layers of the nuclear reaction network. From (a) to (f), figure corresponds to  ${np}$-,  ${nh}$-, ${ph}$-, ${nr}$-, ${pr}$-, and ${hr}$- layers correlation, respectively. Size of circles in each point represents the frequency of correlation, and it's calibration values are inserted.}
\label{degree_correlate}
\end{figure}

\begin{figure}
\vspace{-0.5cm}
\includegraphics[width=9.5cm]{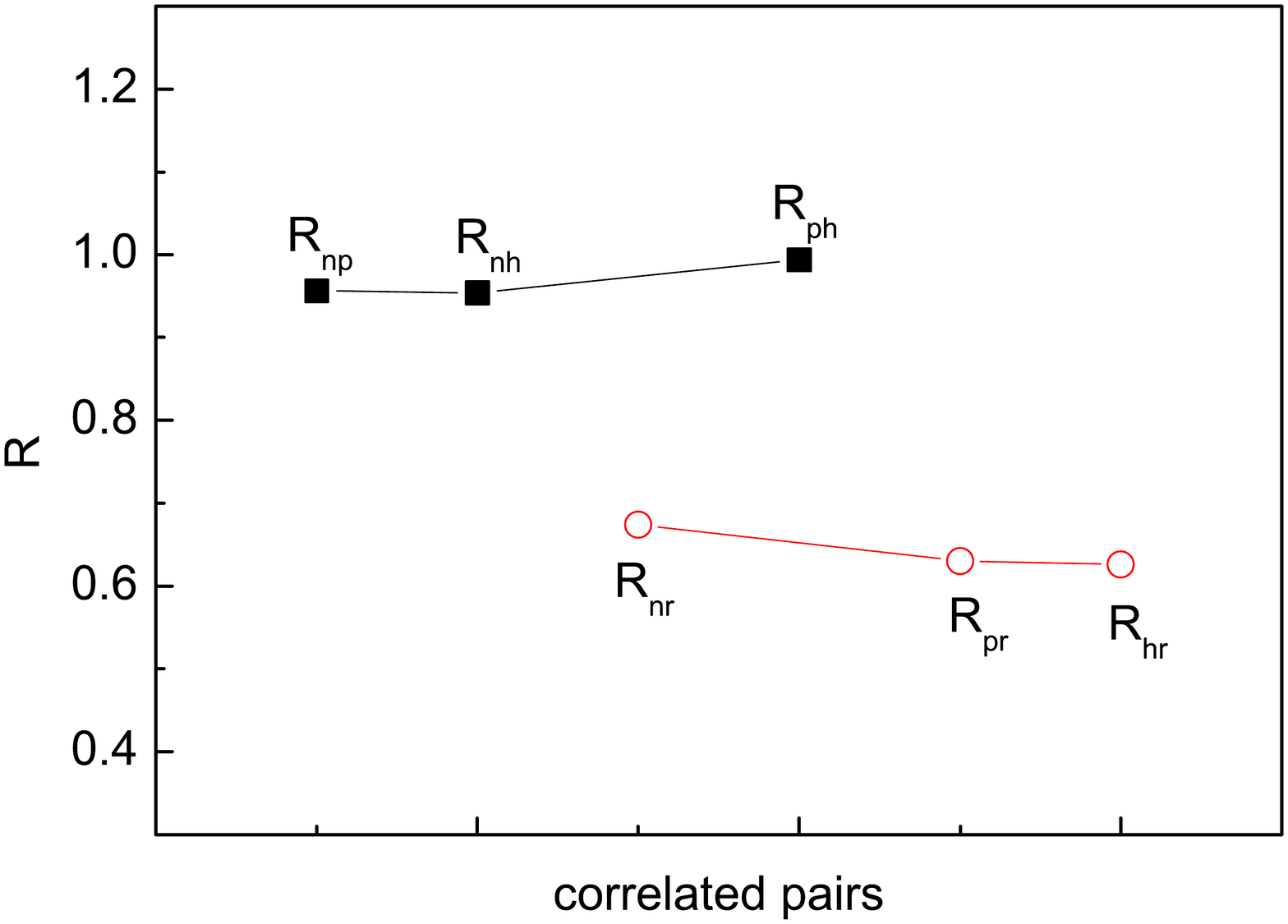}
\vspace{-0.9cm}
\caption{Two-layer degree correlation coefficients among $n$-, $p$-, $h$- and $r$-layers of the nuclear reaction network. $R_{np}$, $R_{nh}$, $R_{ph}$, $R_{nr}$, $R_{pr}$, and $R_{hr}$ represent correlation of $np$-, $nh$-, $ph$-, $nr$-, $pr$- and $hr$-layers, respectively. }
\label{correlate}
\end{figure}

\section{2. Multi-layer nuclear reaction networks}

In the JINA nuclear reaction database, there are 8048 nuclide and 82851 nuclear reactions. Applying the substrate-product method, all nuclei (i.e. nodes) are mapped into a directed un-weighted network which was labeled {by} the edge with three special particles ($n$ for neutron, $p$ for proton or $h$ for $^4$He) as well as
all others ($r$ for {the} remainder).{ For each nucleus (`node' $X$), we can identify the reaction whether it is `in-coming'  (e.g $n$ + $Y \rightarrow X$), or `out-going' ($X + n \rightarrow  Y$). The number of reactions for each nucleus (node) is {represented} by the  `degree' associated with a particular node.} {All nuclear reactions are divided into four classes: the $n$-layer, $p$-layer, $h$-layer and $r$-layer,  each `layer' constituting a type of particle involved in that reaction.} More detailed information can be found in our previous article~\cite{ZL2016-23}. With this definition, the degree value of each nucleus in the landscape can be calculated with each layer of {the} nuclear reaction network, { as shown} in Fig.~\ref{degree}. The degree value of each nucleus is plotted on the $N-Z$ plane with the number of protons $Z$ versus the number $N$ of neutrons. Different degree values are indicated by different {colors}. It indicates that the nuclei in the $n$-layer (a), $p$-layer (b) and $h$-layer (c) have the similar degree value distribution on the $N-Z$ plane, {while such regular attributes exist in this case and a few other values are distributed at the edge of the nuclide map.} {When the number of protons is larger than 82 (one of magic numbers), because of the activity of neutrons, the  distribution of degree values for the nucleus appears different in the $n$-, $p$-, and $h$-layer.} It demonstrates that neutrons also can react with the nucleus in the heavy mass region and result in some nodes with a degree value of 2 in this region, while the $p$ and $h$ layers are not. {However, the $r$-layer (d) has a  distinctive degree value} structure in the $N-Z$ plane. {The main reason of the complexity of the r-layer is the variety of reactions which nuclides can participate, such as $(\gamma,n)$, $(\gamma,p)$, $(\gamma,\alpha)$, $\beta$ decay and so on.} In order to show the degree correlation of each layer, the degree correlation coefficient is defined as
\begin{equation}\label{correlation}
R_{ML} = \frac{\sum_{i}^{N}(K_{i}^{M}-\bar{K}_{i}^{M})(K_{i}^{L}-\bar{K}_{i}^{L})}{\sqrt{\sum_{i}^{N}(K_{i}^{M}-\bar{K}_{i}^{M})^{2}}\sqrt{\sum_{i}^{N}(K_{i}^{L}-\bar{K}_{i}^{L})^{2}}},
\end{equation}
where{,} the $K_{i}^{M}$ and $K_{i}^{L}$ represent the degree of the $i^{th}$ nucleus in the $M$ layer and $L$ layer, respectively. The $\bar{K}^{M}$ and $\bar{K}^{L}$ are the mean degree of all nuclei in the $M$ layer and $L$ layer, respectively. Here the value of $N$ is the number of all nuclides, i.e. 8048, in the nuclear chart.

Fig.~\ref{degree_correlate} and Fig.~\ref{correlate} represent the  degree-degree correlations and their correlation coefficients, respectively,{ for every two layers among $n$-, $p$-, $h$- and $r$-layers of the nuclear reaction network}. {The degree-degree correlations  of $np$-, $nh$- and $ph$-layers are found to be positive, generally displaying almost a linear behavior with some scatterings. }  Furthermore, if we see their degree correlation coefficients $R_{ML}$, i.e. $R_{np}$, $R_{nh}$, and $R_{ph}$,  in Fig.~\ref{correlate}, they are actually very close to 1. Note that the size of circles in each correlated degree represents the frequency  of layer-layer correlation  in Fig.~\ref{degree_correlate}, from which  the upper panels display the degree distributions of $n$-, $p$-, and $h$-layers are basically bimodal, which is very different from other typical degree distributions like either power-law or Poisson types.
{The} degree-degree correlations  of the $r$-layer with $n$-, $p$-, and $h$-layers also display general positive correlation, however, they seem more diverse in comparison with the correlations among $n$-, $p$-, and $h$-layers. Actually, the degree correlation coefficients of $R_{nr}$, $R_{pr}$ and $R_{hr}$ are only about 0.6. {From the degree distributions as shown in Fig.~\ref{degree},  the nuclides in the JINA database have {a} similar structure in the $n$-, $p$-, and $h$- layers of the reaction network.} {However, a relatively complex structure in the $r$-layer of the reaction network is observed from the degree correlation plots (Fig.~\ref{degree_correlate}) and  the correlation coefficients plot $R_{ML}$ (Fig.~\ref{correlate})}.

\section { 3. Topological characteristics and motifs}

{In real systems},  repetitive structures often occur~\cite{RM2002-25,NH2013-26}, such as the regular temporal patterns in the time-varying systems~\cite{LK2013-27,LK2011-28}, the molecular structures in organisms~\cite{JCC2012-18} and even {in the functions or subroutines} repeatedly called in computer programs~\cite{UA2007-30}. {The reaction system in the present scenario also has  repetitive structures. The degree of  each nucleus in the $n$, $p$, and $h$ layers has a regular value except for the boundary and light nuclei regions.} {In network science, such} topological feature{s are  represented as motifs}. Even though the numerical and statistical comparisons are made in the degree value and the correlation coefficient of each layer, the structure of the motif requires a more detailed description, {considering not only} the number of linked edges but also the various possible connections which are examined to distinguish the reaction paths.

The definition of the motif in complex network science is patterns of interconnections occurring in complex networks at numbers that are significantly higher than those in randomized networks \cite{RM2002-25}.
 In general, the ratio of the number of one motif in the real system to that of the motif in the random system, which has the same size, the same number of connected edges and the same degree values of nodes with the real system, is the score of the motif represented as $S_{i}$~\cite{UA2007-30},
$ S_{i} = \frac{O_{i}^{(real)}-\langle O_{i}^{(rand)}\rangle}{\sigma_{i}^{(rand)}}$,
where $O_{i}^{(real)}$ and $\langle O_{i}^{(rand)}\rangle$ are the occurrence number of the $i^{th}$ motif in a real network and the mean occurrence number in a random network, respectively. $\sigma_{i}^{(rand)}$ is the variance of the $i$-th motif in a random network. Of course, different networks have different motif distribution{s}~\cite{UA2007-30}.

\begin{table}
\caption{{Five types of three-node motifs which have the highest statistics in the nuclear reaction network.}}
\begin{tabular}{cccccc}
\hline
\toprule
motif index (id) & 38& 46& 108& 110& 238\\
\hline
\midrule
frequency (unit:$10^4$)& 0.07& 0.66& 0.71& 0.59& 3.87\\
\bottomrule
\hline
\end{tabular}
\end{table}

\begin{table}
\caption{{Seven types of four-node motifs which have the highest statistics in the nuclear reaction network.}}
\begin{tabular}{cccccccc}
\hline
\toprule
motif index (id)& 922& 5064& 5068& 5086& 13150& 13260& 13278\\
\hline
\midrule
frequency (unit:$10^4$)& 1.62& 4.13& 3.37& 2.19& 2.05& 4.24& 11.27\\
\hline
\bottomrule
\end{tabular}
\end{table}

\begin{figure}[htbp]
\small
\centering
\includegraphics[width=9.0cm]{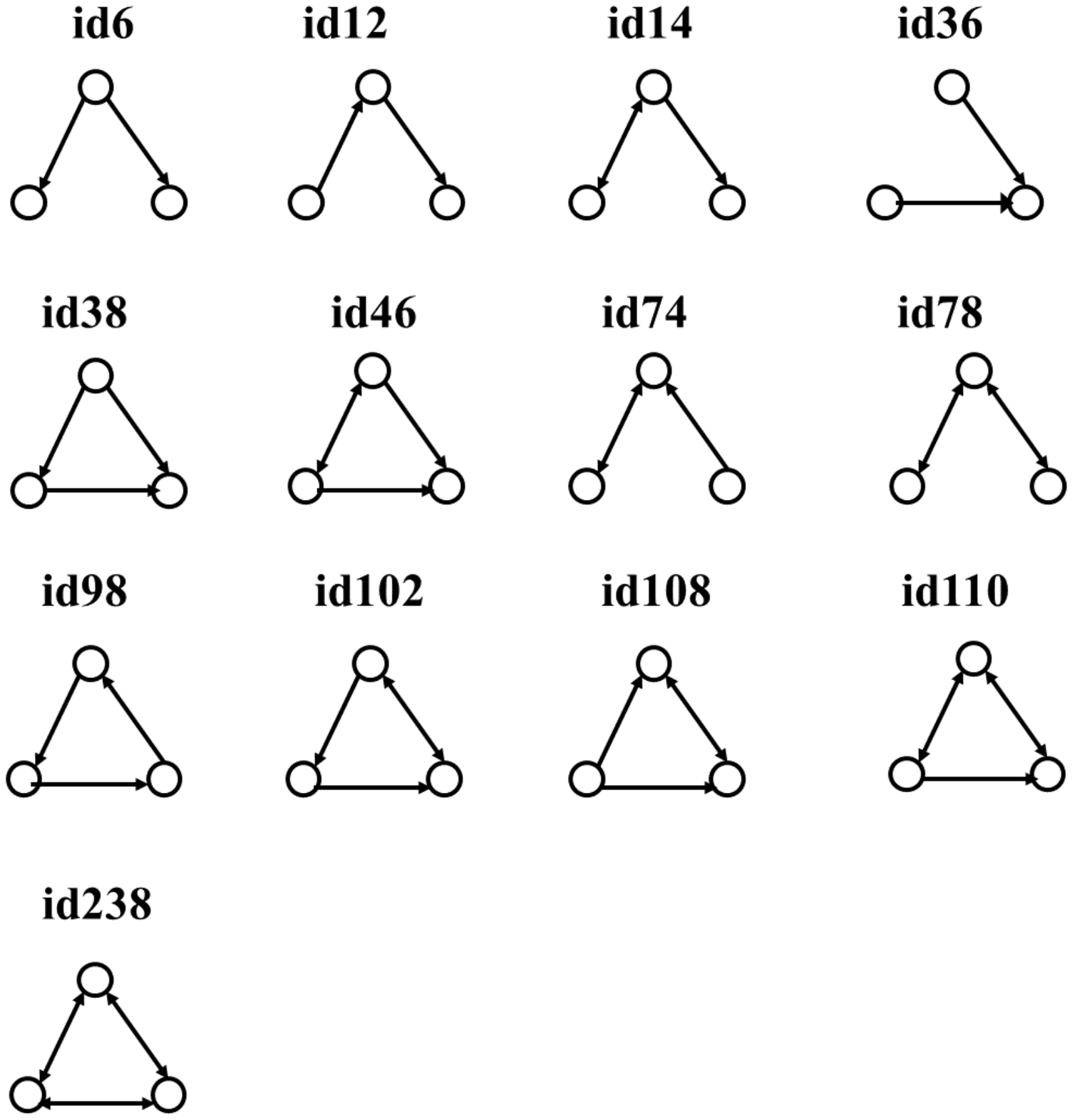}
\vspace{-1.0cm}
\caption{Thirteen types of three-node motifs in the directed network. For directional {single-layer} network, we can sort 13 types of three-node motifs according to various combination of the directional edges among three nodes. Among them, motifs id 38, 46, 108, 110 and 238 occur with high statistics, of which id 238 has the highest frequency. These high statistical motifs reflect that most nodes could be reciprocally transformed and the loop structure is very rich in {the} nuclear reaction network.   }
\label{three-motif}
\end{figure}

\begin{figure}[htbp]
\small
\centering
\includegraphics[width=9.0cm]{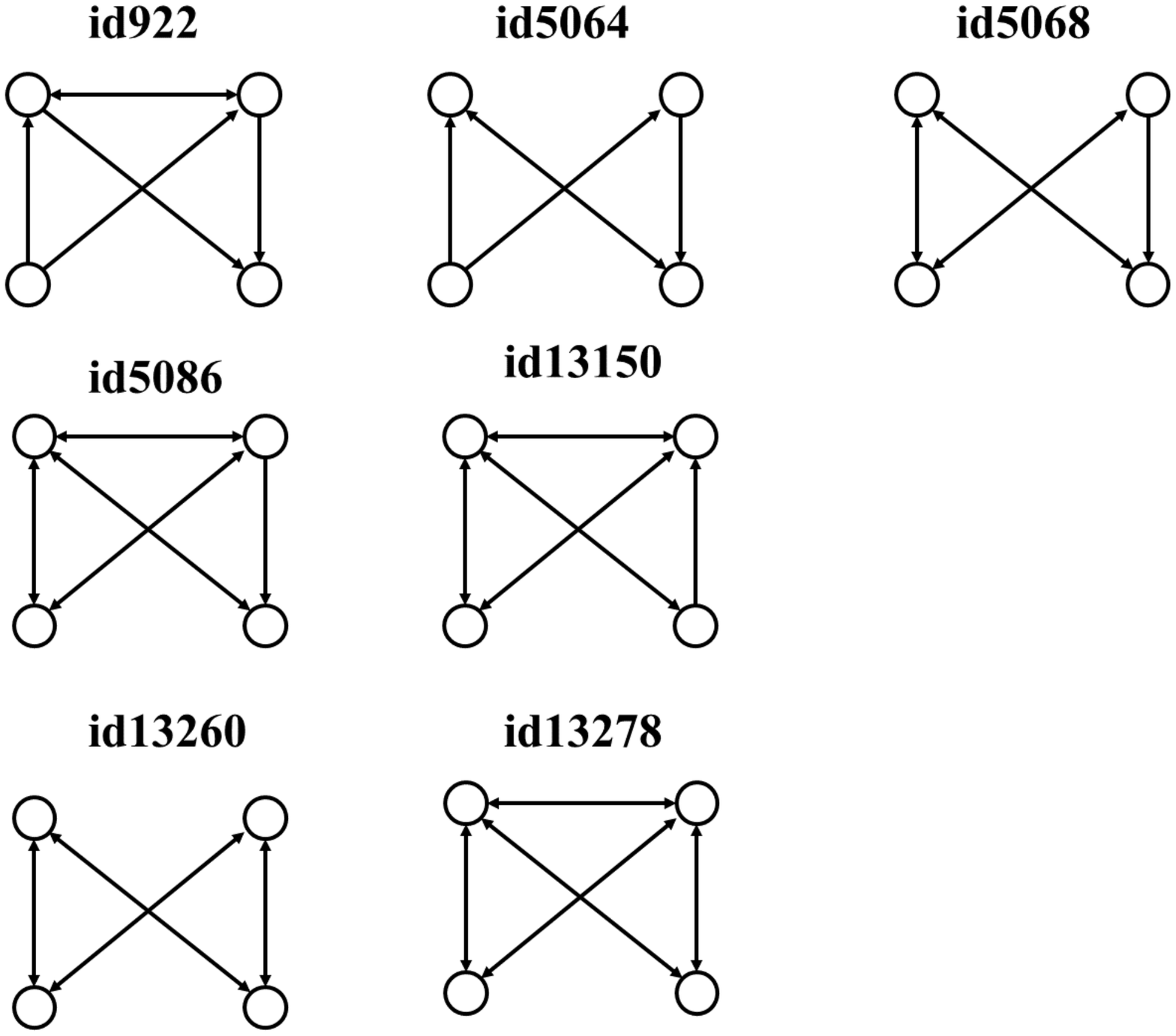}
\vspace{-1.5cm}
\caption{Seven types of four-node motifs which have the highest statistics  in the directed network.
For directional single layer network, we can sort 199 types of 4-node motifs according to various combination of the directional edges among four nodes. Here we just show motif id 922, 5064, 5068, 5086, 13150, 13260 and 13278, which are most frequent 4-node motifs. Id 13278 is the highest statistical motif, which  is  a bidirectionally  connected structure except for absence of a pair of nodes, and it can be seen as a superposition of the above three-node motif id 238 as shown in Fig.~\ref{three-motif}.  }
\label{four-motif}
\end{figure}

The `mfinder'~\cite{NK2002-31}, `MAVisto'~\cite{FS2005-32} and `FANMOD'~\cite{SW2006-33} are the common motif statistical softwares. Here we use the `mfinder' to analyze the motif structure of the nuclear reaction network, which can explore four or more nodes motifs in real networks. According to various combination in the origin of edges with nodes and directions, there are 13 three-node motifs and 199 four-node motifs in a directed network.  Among these, five types of three-node motifs and seven types of four-node motifs have  the highest frequencies in the whole nuclear reaction networks which are shown in Table I and Table II where the {frequencies} of the motifs are given. It can be seen that the number of motifs in the nuclear reaction network is very large{, with a frequency above $10^4$.} The motif type id238 and id13278 are the richest among all three-node and  four-node motifs, respectively. The detailed structure of 13 types of three-node motifs and {seven kinds of four-node} motifs  are shown in Fig.~\ref{three-motif} and Fig.~\ref{four-motif}, respectively. In Fig.~\ref{three-motif}, motif index id 38,  46, 108, 110 and 238 are of highest statistics as listed in Table. I, of which motif id238 is a bidirectional fully connected structure, indicating that most nodes can be transformed into each other in the nuclear reaction network, {e.g.} nucleon absorption and decay.  In Fig.~\ref{four-motif}, motif id13278 is also a bidirectional and has a almost fully connected structure except for the edge absence of a pair of nodes, and it can be seen as a superposition of the above three-node motif id238.  The {absence} of {an} edge between a pair of nodes means that there is no direct reaction path between these two nuclei (nodes), and the distance between two nodes is at least 2 or 3.

\section{4. Discussion}

{It is difficult to make a detailed analysis of the nuclear reaction features using the combination of degree value of each nucleus (in the nuclear landscape) for four layers of the reaction network  and their traditional motif structures.} The nodes such as disseminators, relayers and community cores often have typical neighboring connection patterns so that it is difficult to define the regularity of a structure based on the same degree of each node~\cite{LDF2009-34,LDF2009-35}. {Therefore, a single node motif} has been proposed to analyze the characteristics of each nuclear reaction. In this context,  with the method of regarding each nucleus as a motif, the nuclear reaction paths of each nucleus in the multi-layer nuclear reaction networks have been obtained by learning their in-coming direction as well as out-going direction degrees.

\begin{figure*}[htbp]
\includegraphics[width=18.0cm]{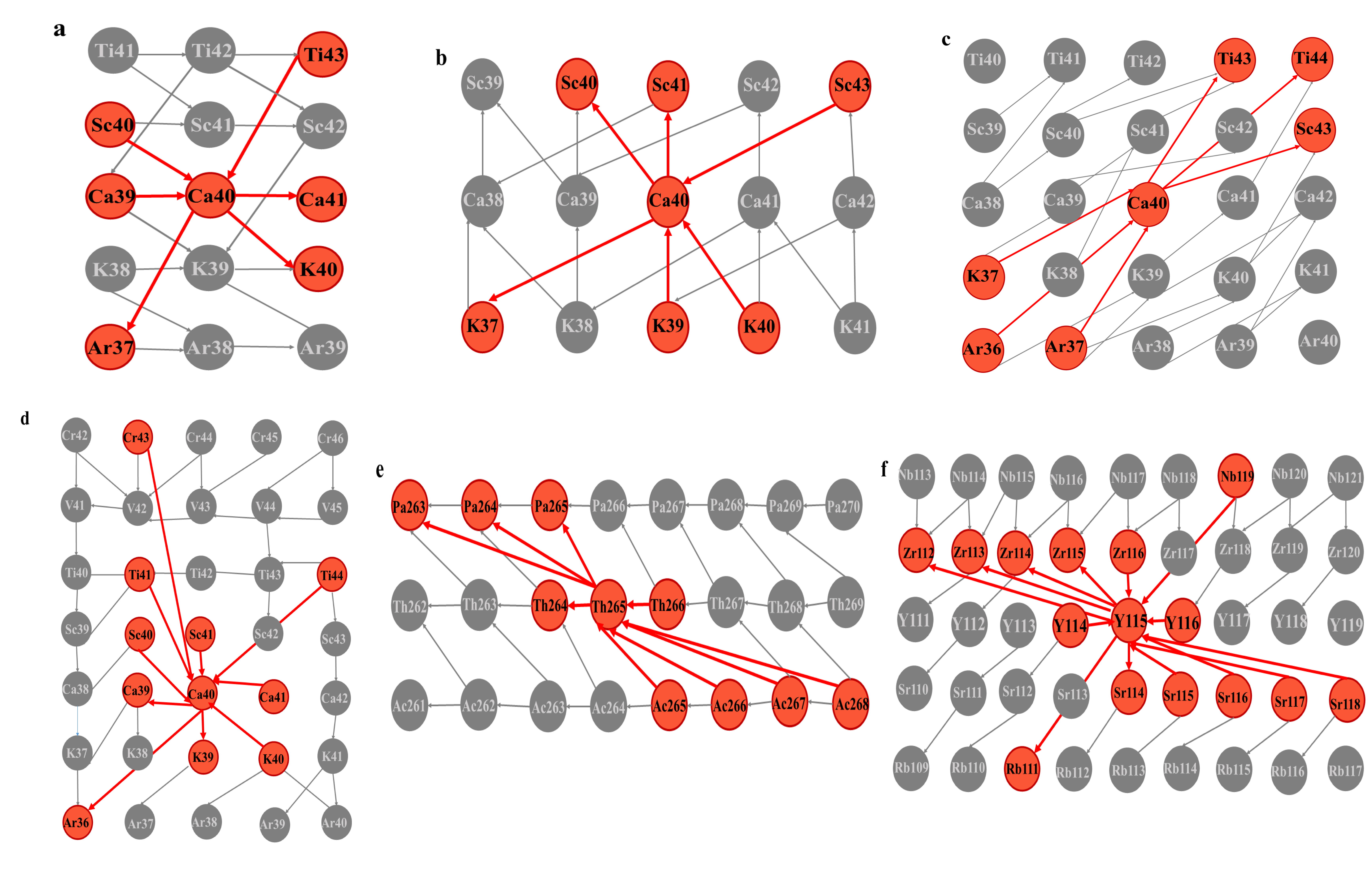}
\caption{ Examples of typical topological structures of motifs in each layer of the nuclear reaction networks. (a), (b) and (c) represents the $n$-layer, $p$-layer and $h$-layer, respectively.    (d), (e) and (f) represents some typical motifs structures of the $r$-layer.  The examples of central nodes are $^{40}_{20}$Ca for (a) - (d), $^{265}_{80}$Th for (e) and $^{115}_{39}$Y (f).
Even though the value of degrees for $n$-, $p$-{,} and $h$-layer (a, b, c) single-node motifs are the same (i.e. $k$ = 6), the effects on the change of proton and neutron number of nuclei are different for different layers, which leads to different topological structure of $n$-, $p$- and $h$-layer motifs. (d) is a typical $r$-layer motif for proton-rich nuclides, while (e) and (f) represent two major motifs for neutron-rich nuclides.
In each plot, the red line with in-coming or out-going arrow illustrates linked in-coming or out-going edge. }
\label{signal-motif}
\end{figure*}

From our analysis, we display  typical structures of motifs of  nuclides in the $n$-, $p$-, $h$-layers and the $r$-layer  in Fig.~\ref{signal-motif}. In the $n$-layer, the nucleus located in ($N,Z$) has in-coming edges from the nuclides located in ($N-1,Z$), ($N-1,Z+1$) and ($N+1,Z+2$), and out-going edges to the nuclides located in ($N+1,Z$), ($N+1,Z-1$) and ($N-1,Z-2$), which correspond to the reactions of ($n$,$\gamma$), ($n,p$) and ($n$,$\alpha$), respectively. However, for nuclide with $Z\geq$ 82 (i.e. Pb), this kind of $n$-layer motif vanishes and replaced by a simpler horizontal structure, i.e. from ($N-1,Z$) to ($N,Z$), then to ($N+1,Z$).
In the $p$-layer, the nucleus located in ($N,Z$) has in-coming edges from the nuclides located in ($N+2,Z+1$), ($N+1,Z-1$) and ($N,Z-1$), and out-going edges to the nuclides located in ($N-2,Z-1$), ($N-1,Z+1$) and ($N,Z+1$), which correspond to the reactions of ($p$,$\alpha$), ($p,n$) and ($p$,$\gamma$), respectively. However, as $Z\geq 83$ (i.e. Bi), this type of $p$-layer motif disappears, and there is no $p$-edge as $Z >84$.
In the $h$-layer, the nucleus located in ($N,Z$) has in-coming edges from the nuclides located in ($N-1,Z-2$), ($N-2,Z-2$) and ($N-2,Z-1$), and out-going edges to the nuclides located in ($N+1,Z+2$), ($N+2,Z+2$) and ($N+2,Z+1$), which correspond to the reactions of ($\alpha$,$n$), ($\alpha$,$\gamma$) and ($\alpha$,$p$), respectively. However, as $Z\geq 81$ (i.e. Tl), this type of $h$-layer motif disappears, and there is no $h$-edge as $Z > 85$.
Viewing from the degree value of the above single-node motif in the three-layer networks, we discover that even though they have the same value {of} degree, the influences of reactions involving neutron, proton, and $\alpha$-particle on the change of neutron and proton numbers { are} quite different due to the nuclear reaction laws, which result in different topological structures of the motifs.

For the complex $r$-layer network, the topological structure of the single-node motif is determined by  reactions of ($\gamma$,$n$), ($\gamma$,$p$), ($\gamma$,$\alpha$), $\beta^{+}$ decay, $\beta^{-}$ decay and so on. For the neutron-rich nuclides, typical  motifs  are shown in Fig.~\ref{signal-motif}(e) and (f). For nuclides with $Z\geq 84$ (i.e. Po), their motif structure will transfer the type (f) for lighter nuclides to the type (e) for heavier nuclides. If proton number will continuously increase, the motif structure in this region will reduce to a type of motif structure with both in-coming degree and out-going degree equal to 3.
For nucleus located in ($N,Z$), its  in-coming edges come from the nuclides located in ($N+1,Z$), ($N+2,Z-1$) and ($N+1,Z-1$), and its out-going edges go to  the nuclides located in ($N-1,Z$), ($N-2,Z+1$) and ($N-1,Z+1$).
For  the proton-rich nuclides, typical motif structure is shown in Fig.~\ref{signal-motif}(d), which is essentially a mirror about $Z$ = $N$  of the motif structure in neutron-rich region, reflecting somehow symmetric feature between neutrons and protons.
In addition, there {is} another important information on stable nuclides in $r$-layer motifs, i.e. comparing with the unstable nuclides, the stable nuclides have one edge difference with reverse direction. If we take Pb as an example, we can see the direction of the edge between stable nucleus $^{207}$Pb and its neighbor ($N+1,Z-1$) is opposite to that of the edge between unstable nucleus $^{205}$Pb and its neighbor ($N+1,Z-1$). This difference between stable and unstable nuclides can also explain why the condition of stable nuclides is that the difference of degree in $r$-layer is two, i.e., $K_{i,\mathrm{in}}^{[r]}-K_{i,\mathrm{out}}^{[r]}=2$,  which was proposed in our previous work \cite{ZL2016-23} .

All the above typical motifs can be continuously repeated and overlapped in the specific $r$-layer region or almost in whole $n$-, $p$- and $h$-layers, but only some nodes in the $r$-layer have long-range edges.  Also, we note that  all topological structures revealing motifs of each layer in the multi-layer nuclear reaction network tend to fade away when the proton number is larger than the magic number of $Z$ = 82 or equal to its neighboring value. This, in turn, indicates that the topological structure of nuclide in this region becomes simpler or tends to disappear. This seems to be a global feature for the entire nuclear landscape.

\section {5. Summary}

In summary, we construct a directed un-weighted multi-layer nuclear reaction network, which consists of all nuclides and reactions in the JINA REACLIB database, to investigate topological features and motifs structure of the nuclear reactions.  Considering there are four types of edges, namely neutron, proton, $^4$He and remainder, we present four-layer nuclear reaction networks,  which correspond to $n$-layer, $p$-layer, $h$-layer and $r$-layer.
From topological structures of the nuclear reaction networks, we find that the nuclear reaction networks of the $n$-, $p$- and $h$-layers have a regular structure and they {can} be regarded as  the repetitive superposition of  motifs which have spatial features. The $r$-layer network has multiple motifs, but they also belong to the same category. For the motifs structure in all different layers, they fade away or change with the protons number above the magic number 82 or around.
By contrasting with motifs structure of  typical stable nuclides and those of its neighboring unstable nuclides in the r-layer, we find that in most cases the only difference  is the direction of a certain edge turns reverse, i.e. from in-coming to out-going or vice versa, which also explains why the degree difference between in-coming  and out-going in the $r$-layer for the stable nuclides  is two \cite{ZL2016-23}. Moreover, it is found that the degree-degree correlation are strong for $np$-, $nh$-, and $ph$-layers, generally displaying almost linear behavior with some scatterings.   Their degree correlation coefficients $R_{ML}$  are very close to 1. However, for degree-degree correlations  of the $r$-layer with $n$-, $p$-, and $h$-layers, they seem weakly correlated, and the degree correlation coefficients are  only about 0.6, which are also consistent with different motif structures of the $n$-, $p$-, $h$-layers from the $r$-layer. {This study presents a unique understanding of thermonuclear reactions} in astro-nuclear process.  {A further study considering the $r$-process, and the other nuclear reaction processes involving all the nuclides may be beneficial.}

{\it Acknowledgments.--}
This work is supported by the National Natural Science Foundation of China under Contracts Nos. 11890714,1421505, 11875133 and 11075057,  the National Key R$\&$D Program of China under Grant No. 2018YFB2101302.

\end{document}